\documentclass[manuscript,screen , noacm]{acmart}
\AtBeginDocument{%
  \providecommand\BibTeX{{%
    \normalfont B\kern-0.5em{\scshape i\kern-0.25em b}\kern-0.8em\TeX}}}

\usepackage[showdeletions]{color-edits}
\usepackage{xspace}
\usepackage{color, colortbl}
\usepackage{xcolor}
\usepackage{enumitem}
\usepackage[singlelinecheck=false]{caption}
\usepackage{tabularx}  
\usepackage{multirow}
\usepackage{multicol}
\usepackage{threeparttable}
\usepackage{float}
\usepackage[bottom]{footmisc}
\newcolumntype{P}[1]{>{\centering\arraybackslash}p{#1}}
\newcolumntype{R}[1]{>{\raggedleft\let\newline\\\arraybackslash\hspace{0pt}}m{#1}}
\newcolumntype{L}[1]{>{\raggedright\let\newline\\\arraybackslash\hspace{0pt}}m{#1}}

\definecolor{dawnblue}{rgb}{0.84, 0.92, 1.0}

\usepackage[capitalize,noabbrev]{cleveref}

\setlist[itemize]{align=parleft,left=0.5em..1.5em}

\newlist{req}{enumerate}{2}
\setlist[req,1]{label=RQ \arabic*:,ref= \textbf{\arabic*}, leftmargin=*}

\newlist{hyp}{enumerate}{2}
\setlist[hyp,1]{before=\itshape,font=\itshape, label=Hypothesis \arabic*:,ref= \arabic*, leftmargin=*}
\setlist[hyp,2]{before=\itshape,font=\itshape, label=Hypothesis \arabic*:,ref= \arabic*, leftmargin=*}

\addauthor{jh}{blue} 
\addauthor{ps}{blue} 
\addauthor{ph}{blue} 
\addauthor{dm}{blue} 
\addauthor{nk}{blue} 

\begin{document}

\title[The Effect of Contextual Information]{Human Delegation Behavior in Human-AI Collaboration: \\
The Effect of Contextual Information}


\author{Philipp Spitzer}
\email{philipp.spitzer@kit.edu}
\orcid{0000-0002-9378-0872}
\authornote{Both authors contributed equally to this research.}
\affiliation{%
  \institution{Karlsruhe Institute of Technology}
  \country{Germany}
}

\author{Joshua Holstein}
\email{Joshua.Holstein@kit.edu}
\orcid{0009-0005-3885-8365}
\authornotemark[1]
\affiliation{%
  \institution{Karlsruhe Institute of Technology}
  \country{Germany}
}

\author{Patrick Hemmer}
\email{Patrick.Hemmer@kit.edu}
\orcid{0000-0003-3159-9602}
\affiliation{%
  \institution{Karlsruhe Institute of Technology}
  \country{Germany}
}

\author{Michael Vössing}
\email{Michael.Voessing@kit.edu}
\orcid{0000-0002-7722-6142}
\affiliation{%
  \institution{Karlsruhe Institute of Technology}
  \country{Germany}
}

\author{Niklas Kühl}
\email{kuehl@uni-bayreuth.de}
\orcid{0000-0001-6750-0876}
\affiliation{%
  \institution{University of Bayreuth}
  \country{Germany}
}

\author{Dominik Martin}
\email{Dominik.Martin@kit.edu}
\orcid{0000-0002-2166-3183}
\affiliation{%
  \institution{Karlsruhe Institute of Technology}
  \country{Germany}
}

\author{Gerhard Satzger}
\email{Gerhard.Satzger@kit.edu}
\orcid{0000-0001-8731-654X}
\affiliation{%
  \institution{Karlsruhe Institute of Technology}
  \country{Germany}
}

\renewcommand{\shortauthors}{Spitzer et al.}

\begin{abstract}

The integration of artificial intelligence (AI) into human decision-making processes at the workplace presents both opportunities and challenges. One promising approach to leverage existing complementary capabilities is allowing humans to delegate individual instances of decision tasks to AI. However, enabling humans to delegate instances effectively requires them to assess several factors.
One key factor is the analysis of both their own capabilities and those of the AI in the context of the given task.
In this work, we conduct a behavioral study to explore the effects of providing contextual information to support this delegation decision. Specifically, we investigate how contextual information about the AI and the task domain influence humans' delegation decisions to an AI and their impact on the human-AI team performance. Our findings reveal that access to contextual information significantly improves human-AI team performance in delegation settings. Finally, we show that the delegation behavior changes with the different types of contextual information. Overall, this research advances the understanding of computer-supported, collaborative work and provides actionable insights for designing more effective collaborative systems.

\end{abstract}





\maketitle

\section{Introduction}
\label{sec: Introduction}

The rapid growth of artificial intelligence (AI) is fundamentally reshaping a wide range of domains, including healthcare, manufacturing, and finance \cite{Raees2023, Gmeiner2023, Liu20222, yu_artificial_2018, spitzer2024transferring, holstein2025designing}. In many fields, AI can assist humans in their decision-making by providing recommendations and even solving specific task instances autonomously. Despite these capabilities, effective collaboration with human decision-makers is typically required, posing significant design challenges for computer-supported cooperative work (CSCW) systems. This necessity arises from two main considerations: the advantage of leveraging the complementary capabilities of humans and AI to achieve superior team performance \cite{hemmer2022forming, hemmer2022effect} and regulatory frameworks, such as the European AI Act, that mandate human oversight to ensure accountability and transparency \citep{zanzotto2019human, european_commision_european_2023}.

Despite the remarkable capabilities of AI in specialized tasks, its integration into human workflows remains a complex challenge \cite{Cabitza2023, morrison2024impact}: the effectiveness of human-AI collaboration---the interaction between humans and AI to solve tasks and make decisions collaboratively---is fundamentally linked to how people perceive, understand, and ultimately use these systems \cite{LOGG201990, Cabitza2023, SUTTON2023100626, chong2022human}. Among the various ways to integrate AI in CSCW workflows, human delegation evolves as a critical form. Here, humans can assign task instances to an AI when they believe the AI can perform the tasks more effectively or efficiently than they can themselves \cite{fugener2022cognitive, taudien2022calibrating, hemmer2023human, taudien2022effect}. However, designing systems that effectively manage this delegation of tasks between humans and AI remains challenging.

To address this, it is important to understand the factors that influence human decision-making when delegating tasks to an AI. Recent research suggests that two factors are essential in this context: the perceived difficulty of the instance \cite{spitzer2023perception} and the human perception of efficacy \cite{hemmer2023human, pinski2023ai, fugener2022cognitive}---both an individual's belief in their ability to successfully execute tasks (self-efficacy) and the ability attributed to the AI (AI efficacy). While high self-efficacy may reduce the preference to delegate tasks to AI, low self-efficacy may lead to over-delegation \cite{taudien2022calibrating, fugener2022cognitive}, bypassing opportunities to apply human expertise and judgment \cite{Cabitza2023}. Similarly, the perception of the AI's efficacy plays a critical role \cite{pinski2023ai}: discrepancies in the perceived capabilities of AI, whether over- or underestimation, may cause ineffective delegation decisions.

Among the factors shaping these perceptions, contextual information is paramount \cite{kawakami2022improving, holstein2023toward}.
Within human-AI collaboration, contextual information refers to any additional information beyond simple AI predictions that is provided to support humans in making better decisions for a given task. Examples of contextual information include supplemental information not available to the AI \cite{hemmer2022effect}, general information to increase AI literacy \cite{pinski2023ai}, explainable AI \cite{schemmer2023towards}, and AI uncertainties \cite{taudien2022calibrating}. 
While existing research has extensively examined how explainable AI and AI uncertainties influence human-AI decision-making and human delegation specifically, research lacks an understanding of how contextual information about the underlying data and the AI team partner influence humans' delegation behavior. Further, despite the crucial role of human delegation, only a few studies investigate the role of human factors that shape the delegator's decisions on whether or not to delegate a task instance to an AI.

\begin{figure}[!ht]
\centering
\includegraphics[width=1.0\textwidth]{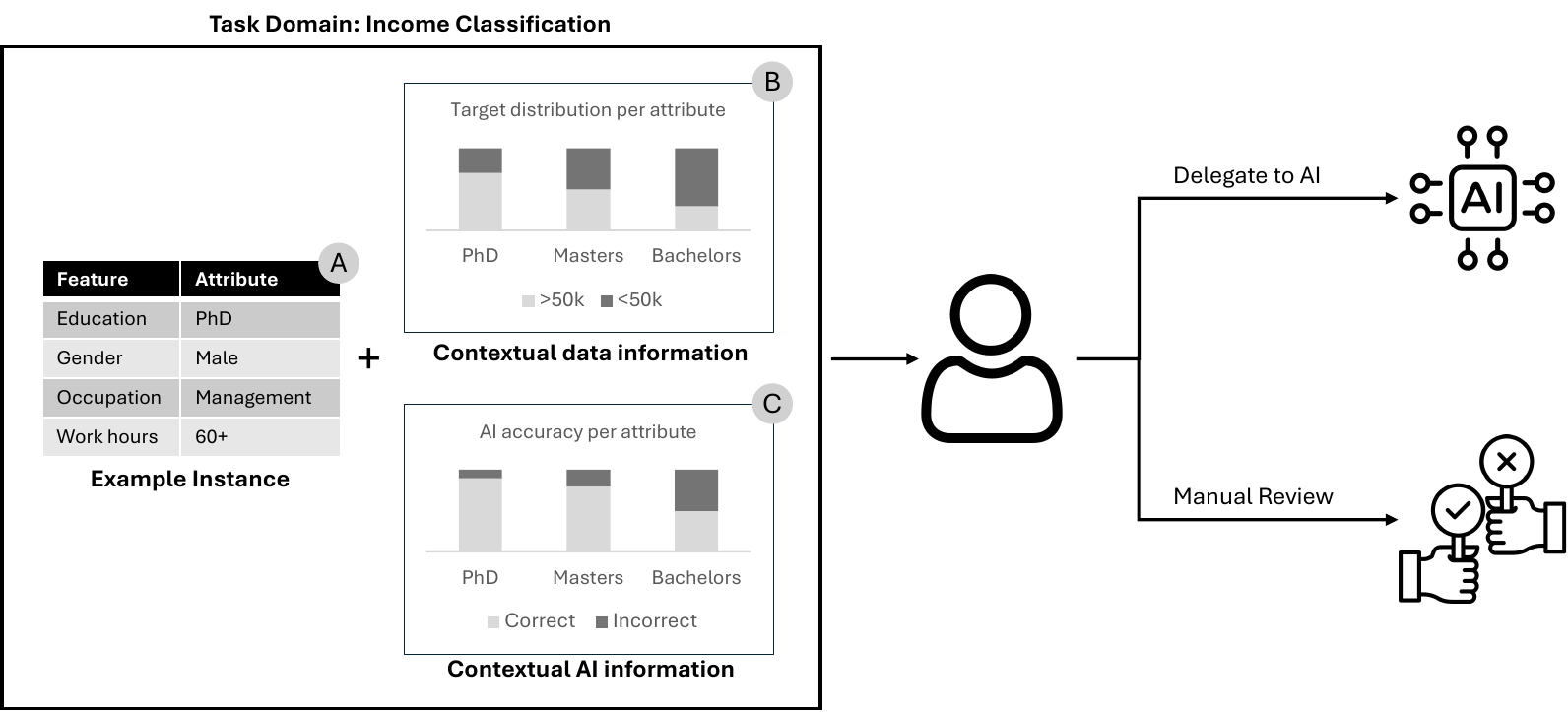}
\caption{In our human-AI delegation study, participants receive an instance (A) and two types of contextual information: (B) The distribution of a specific attribute within the data and (C) the AI's accuracy in relation to this attribute. This information supports participants in making informed decisions about whether to handle each instance manually or delegate it to the AI, optimizing the balance between human expertise and AI capabilities. During the study, the distributions of all features of an instance are available.}
\label{fig:teaser}
\end{figure}

Consequently, we focus in this study on two types of contextual information: (1) supplementary information on the distribution of data regarding a target variable (contextual data information) and (2) supplementary information on AI’s accuracy across different attributes concerning a target variable (contextual AI information). As these types of contextual information have the potential to provide human delegators with a more holistic and accurate understanding of the task, data, and their AI team partner, we aim to help them make more informed decisions about when and how to delegate tasks to the AI (see \cref{fig:teaser}).

Providing contextual information to human delegators may help them assess the difficulty of individual tasks more accurately, which in turn can influence their perceptions of self- and AI efficacy and ultimately affect delegation behaviors and human-AI team performance. By calibrating these perceptions, we aim to improve delegation decisions. However, existing research suggests that more information does not necessarily lead to better performance and may even worsen it \cite{dietzmann2022artificial, Eppler2004}. This phenomenon is often attributed to cognitive overload, where the information presented exceeds an individual's processing capacity, causing them to overlook or misinterpret relevant data \cite{sweller1994cognitive}. Therefore, it is essential to carefully design the presentation and format of contextual information to avoid cognitive overload.

Consequently, this study examines whether and how the provision of contextual data and AI information affects human-AI team performance. Specifically, it examines how different types of information affect human delegation behavior, identifies the human factors that mediate these effects, and assesses their overall impact on human-AI team performance. Thus, we ask the following research questions

\begin{req}[labelindent=1em, labelwidth=0em, label=\textbf{RQ\arabic*}:, ref=\arabic*]

    \item How does providing varying types of contextual information impact human-AI team performance in human delegation? \label{rq1}
    \item What factors mediate the impact of contextual information on the human-AI team performance? \label{rq2} 
    \item How do humans’ perceived efficacy and humans’ perceived difficulty correlate to the delegation behavior for varying types of contextual information? \label{rq3}

\end{req}

To answer our research questions, we conduct a between-subjects study (n=240) with four treatments, each receiving different forms of contextual information. In our experiment, we measure the impact of these contextual information types on delegation behavior and human-AI collaboration performance for an income prediction task. We adopt established metrics from previous research that asses the perceived instance difficulty \cite{spitzer2023perception}, self-efficacy \cite{hemmer2023human}, and the AI efficacy \cite{pinski2023ai}.

The contribution of our work is threefold: (1) we show that providing contextual information based on either data distribution or AI performance does not significantly increase human-AI team performance. Instead, only when provided with both types of contextual information, the human-AI team performance can be increased considerably. (2) Through a mediation analysis, we identify underlying human factors that impact the human-AI team performance within delegation settings. When humans have access to both types of contextual information, their perception of the AI efficacy and instance difficulty mediate the impact on the human-AI team performance. (3) We shed light on how humans' perceived efficacy and their perceived difficulty influence their delegation decisions for varying types of contextual information. We observe that they delegate instances more often when they find them difficult. Additionally, we see that providing contextual information allows us to shape humans' delegation decisions, i.e., contextual data information leads to less delegation while contextual AI information results in more instances delegated to the AI.

The remainder of this work is structured as follows: first, we outline related literature on human-AI delegation (\Cref{sec: Related Work},) before describing our theoretical development (\Cref{theoretical_section}) and methodology for our empirical study (\Cref{methodology}). Then, we present the results of our work (\Cref{results_section}). After that, we reveal the implications and limitations of our study (\Cref{sec: Limitations}) and conclude our work (\Cref{conclusion_section}).

\section{Related Work}
\label{sec: Related Work}
Exploring different delegation mechanisms and their underlying rationales is at the core of research in human-computer interaction (HCI) and CSCW. Understanding when to delegate task instances to an AI or execute them manually is crucial for effectiveness at the workplace. In this section, we review relevant works focusing on the role of contextual information in human-AI collaborations. We first outline works that investigate the role of contextual information in human-AI collaborations where the AI is integrated as a team member, providing assistance in the form of advice to the decision-maker. Afterward, we outline works that explore the role of contextual information in delegation scenarios, where humans can delegate specific task instances to the AI.

\subsection{Contextual Information in Human-AI Collaboration}
\label{sec: contextinf}
Previous research in HCI and CSCW has explored factors that influence the collaborative behavior between humans and AI. A key aspect of this interaction is access to contextual information that influences human collaborative behavior. In general, contextual information refers to additional information that supports the human's mental models of an AI or the task domain. 

Previous work investigates which contextual information equips decision-makers the best in the context of AI-assisted decision-making. \citet{Cai2019} investigate the type of information that pathologists require from AI systems in order to integrate them into their work. Among others, they identify precise accuracy information on the data for different target classes as well as additional information about the input data as critical factors. \citet{Yin2019} find that the stated accuracy of the AI influences the participants' decision and their trust in the AI model. Furthermore, \citet{Kawakami2023} show that training decision makers through feedback and additional information about the AI improves their ability to identify instances where the AI is wrong. \citet{Lai2020} investigate the effect of providing guidelines and explanations during onboarding on human performance in working with an AI.

Other works investigate different forms of contextual information to facilitate human-AI collaboration. Different studies explore the effect of providing explanations for AI predictions. Amongst others, \citet{taudien2022calibrating} provide participants with AI uncertainties and information on the AI accuracy to support participants' decisions. The authors show that such contextual information can lead to improved human-AI team performance and trust in AI. Similarly, \citet{ferreira2021human} provide explanations to radiologists as contextual information. The authors argue that access to contextual information is crucial in physicians' decision-making processes.
However, a meta-analysis of \citet{schemmer2022meta} yields mixed results regarding the effectiveness of including explanations with AI predictions, showing that providing explanations often does not lead to improved decision-making.
Other works have studied the effects of providing information about the AI to participants. \citet{pinski2023ai} provide contextual information in the form of general knowledge on AI models, i.e., AI literacy and their underlying principles, and demonstrate their effectiveness in improving human-AI team performance. \citet{Chiang2021} inform participants that AI models can have varying accuracy across different subsets of the data and let users explore the AI's performance for different subsets. \citet{kawakami2022improving} state that humans rely on their understanding of the AI's strengths and weaknesses, but also their extent of contextual information. Thus, humans' mental model of the AI and the given context plays a crucial role. Moving to contextual information on the data, \citet{schemmer2022influence} establish the conceptualization of information asymmetry, stating that to reach complementary team performance, humans' access to different sources of contextual information presents a key factor. Picking up on this theory, \citet{holstein2023toward} investigate the impact of information access of unobservables on humans' performance. The authors show that access to such information can change human interaction behavior. 
Similarly, \citet{agarwal2023combining} empirically analyze how the provision of contextual information and AI predictions impact the diagnosis quality of radiologists. They show that providing AI predictions does not increase the diagnosis quality, but access to contextual information does. 

As previous research shows, contextual information in AI-assisted decision-making is a crucial factor to consider when developing human-AI collaboration scenarios. While many works exist in the context of providing contextual AI information, only a few investigate the effects of additional information on the underlying data despite its importance, as suggested by \citet{Cai2019, holstein2024understanding}.

\subsection{Human Delegation}
\label{sec: Human_Delegation}
In recent years, the field of human-AI collaboration has received significant attention, intending to utilize the synergistic capabilities of both humans and AI. One such form denotes delegation collaborations. Researchers examine extensively two primary forms of this collaboration: AI delegating instances to humans \citep{fugener2022cognitive, hemmer2023human, leitao2022human, bondi2022role, hemmer2022forming, keswani2021towards, mozannar2020consistent, okati2021differentiable, raghu2019algorithmic, wilder2020learning} and humans delegating instances to AI \citep{fugener2022cognitive, lubars2019ask, milewski199W7delegating, taudien2022calibrating, fugener2021will, steffel2018delegating, pinski2023ai}. While both forms hold immense potential, our study is dedicated to investigating the latter and delving into the dynamics and implications of humans entrusting AI systems with single instances. This exploration sheds light on the evolving landscape of human-AI collaboration, emphasizing AI's essential role as a competent and reliable delegate in a broad range of applications. For example, in \citet{lubars2019ask}, the authors investigate factors that influence the human delegation decision. They show that motivation, risk, trust, and difficulty play an essential role in the delegation process. \citet{milewski1997delegating} study the essential constructs for designing collaborative systems while acknowledging the impact of users' mental models on the delegation process. \citet{taudien2022calibrating} explore how humans delegate instances to an AI when they are provided with the AI's uncertainties. \citet{pinski2023ai} explore how AI literacy affects humans' delegation behavior. Similarly, \citet{fuegener2021exploring} analyze the delegation behavior of humans in human-AI collaboration. They cluster participants based on performance, delegation rate, and self-assessment and identify various types of delegation behaviors.

While previous research takes first steps in exploring the factors and partially the role of contextual information in human delegation scenarios, we extend the knowledge in HCI and CSCW by further investigating how different types of contextual information (e.g., contextual AI information and contextual data information) influence human delegation behavior and explore driving human factors that influence this influence of contextual information. By doing so, we build on previous CSCW works \citep{Cai2019, pinski2023ailiteracy, taudien2022calibrating} that have so far evaluated one specific form of contextual information but limited themselves to instance-specific information about the AI (e.g., AI's uncertainty in the prediction). With the advancements in the field to provide users with more helpful information on the AI (e.g., descriptions of AI behavior \citep{cabrera2023improving}), we make use of these developments and explore not only instance-specific contextual information on the AI but on a global level. Next to this extension of prior knowledge in delegation scenarios, we bring in a novel view on human-AI delegation by investigating not only contextual information about the AI but also about the domain in the form of data-specific information. By doing so, we extend the knowledge of the influence of different forms of contextual information on delegation behavior. With prior research only focusing on the provision of local, AI-specific contextual information, we expand the understanding of the role of contextual information in human-AI delegation scenarios. We explore this impact empirically through an online study.
\section{Theoretical Development}
\label{theoretical_section}

In the evolving field of human-AI collaboration, understanding the delegation of tasks from humans to AI involves considering a variety of factors. Drawing on relevant research, this section aims to develop a research model to explore this delegation behavior. To do so, we draw from theories of adjacent research fields in organizational behavior and psychology and synthesize factors that influence human delegation behavior.

While humans typically have a preference for maintaining control over the task \citep{miller1979controllability}, access to contextual information has the potential to modify this behavior \citep{pinski2023ai, hemmer2022effect}. \citet{hemmer2022effect} define the distinct access to information as information asymmetry and show its role in the interaction of humans and AI. Similar studies point out the role of information access and its impact on humans' behavior in human-AI collaboration scenarios \citep{holstein2023toward, pinski2023ailiteracy}. For example, \citet{pinski2023ai} investigate how humans' delegation behavior is impacted by knowledge regarding the strengths and weaknesses of AI in comparison to humans. The authors show that this additional knowledge moderates the effect of humans' appraisal on their delegation behavior. Especially in collaboration scenarios, in which humans possess different levels of domain knowledge, their delegation behavior may differ. For example, financial experts who need to evaluate customers' incomes might possess knowledge of contextual data information (i.e., the share of people with an occupation in professional sports that have an income higher than \$50,000) as they have years of experience in this task domain. Contrarily, in collaboration with an AI, not only humans' general knowledge of AI \citep{pinski2023ai, cardon2023challenges} but also their knowledge about a specific AI, for example, its predictive performance on subsets \citep{cabrera2023improving} or error boundaries \citep{bansal2019beyond}, play a central role which impacts human-AI interaction \citep{bansal2019beyond, schemmer2022influence, mohseni2021multidisciplinary}. We assume in relation to \textbf{RQ \ref{rq2}} that such knowledge influences the human-AI team performance. Thus, we hypothesize:

\begin{hyp}[resume, wide, leftmargin=0cm, labelindent=0pt, labelwidth=0em]
\item Having access to contextual information improves human-AI team performance. 
\label{hyp1}
\end{hyp}

Research in HCI explores the effect of underlying factors that influence human-AI delegation. In the context of AI delegation, \citet{hemmer2023human} reveal self-efficacy to be an underlying mediator of task performance and task satisfaction. Generally, self-efficacy refers to an individual's level of confidence in their ability to perform a given task proficiently \citep{bandura1977self}. In the work of \citet{bandura1997}, the authors explore how individuals' beliefs in their ability impact their performance. The work shows that humans' self-efficacy enhances their accomplishments. Self-efficacy also plays an important role in behavioral research. For instance, \citet{strecher1986role} show that higher levels of self-efficacy lead to better performance in the adoption of health-promoting behaviors. Prior research has highlighted the relevance of self-efficacy in the context of delegation scenarios. For example, experiments in organizational research suggest that supervisors delegating tasks to employees can enhance their psychological empowerment \citep{zhang2017leaders}. Hence, delegation instills in employees a sense of job significance and accountability for the outcomes of their work. Other studies show that self-efficacy exerts influence over the learning process and the extent of effort invested in work tasks \cite{lunenburg2011self}. In a related study, \citet{pinski2023ai} explore the effect of a similar construct (e.g., task appraisal) on the delegation behavior of humans. Task appraisal refers to humans' perceived suitability to solve the task. \citet{pinski2023ai} show in their work that humans' task appraisal influences the human-AI team performance. 
Based on these previous works that show the influence of self-efficacy in delegation settings, we assume that humans' instance-specific efficacy for themselves and the AI impacts their delegation behavior for single instances in a classification task. If a human perceives themselves as suitable for an instance or if a human perceives the AI as suitable for an instance, this will likely impact the human-AI team performance. Thus, in order to answer \textbf{RQ \ref{rq2}}, we hypothesize:

\begin{hyp}[resume, wide, leftmargin=0cm, labelindent=0pt, labelwidth=0em]
\item Humans' instance-specific self-efficacy mediates the effect of contextual information on the human-AI team performance in human-AI collaboration.
\label{hyp2}
\end{hyp}

\begin{hyp}[resume, wide, leftmargin=0cm, labelindent=0pt, labelwidth=0em]
\item Humans' instance-specific AI efficacy mediates the effect of contextual information on the human-AI team performance in human-AI collaboration.
\label{hyp3}
\end{hyp}

Research in psychology studies how task difficulty affects human behavior. For instance, \citet{goldhammer2014time} show that a higher task difficulty leads to more time needed before switching to the next task. In addition, \citet{kukla1974performance} and \citet{scasserra2008influence} show that the perception of task difficulty affects the performance on the task. Related research reveals the impact of instance-specific difficulty perception on humans' delegation behavior \citep{steffel2018delegating}. In their study, \citet{steffel2018delegating} show that humans delegate instances that they perceive as difficult more likely to an intelligent agent. Thus, the difficulty perception of humans influences their delegation behavior. \citet{lubars2019ask} establish a framework of human factors influencing delegation decisions. The authors reveal that the perceived instance difficulty does influence humans' delegation decisions. In a related study, \citet{fugener2022cognitive} analyze humans' difficulty perception of the instance in relation to their delegation behavior. They outline that humans struggle to assess their difficulty perception appropriately, impairing their delegation behavior. In relation to \textbf{RQ \ref{rq2}}, we assume that humans' access to contextual information will likely impact their difficulty perception of instances. Likewise, we assume that their difficulty perception influences the human-AI team performance. Accordingly, we hypothesize:

\begin{hyp}[resume, wide, leftmargin=0cm, labelindent=0pt, labelwidth=0em]
\item Humans' instance-specific difficulty perception mediates the effect of contextual information on the human-AI team performance in human-AI collaboration.
\label{hyp4}
\end{hyp}

\begin{figure}[!htb]
  \centering
  \includegraphics[width=\linewidth,clip]{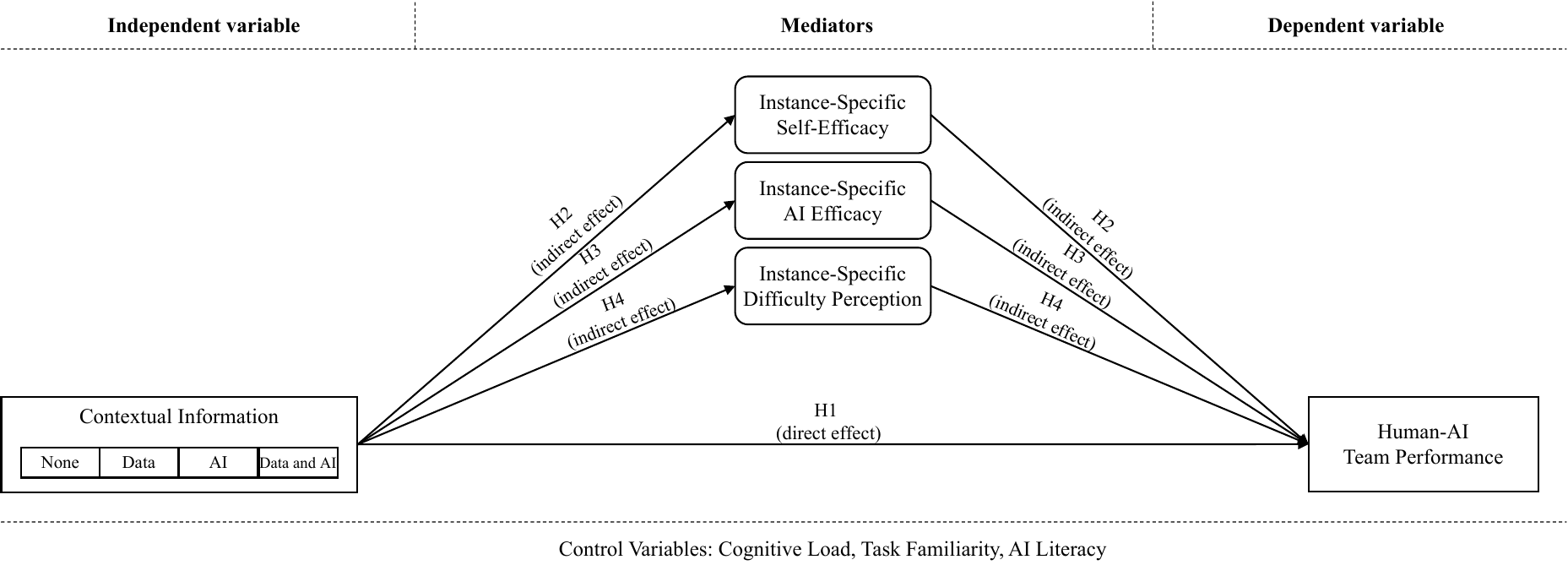}
  \caption{The research model describes the relationship of the independent variable (contextual information), mediators (instance-specific self-efficacy, instance-specific AI efficacy, instance-specific difficulty perception) and the dependent variable (delegation performance).}
  \label{research_model}
  \Description{The graph shows several boxes connected to each other via arrows. The one on the left has four arrows going out and is titled "Contextual Information" with values "None", "Data", "AI" and "Data and AI". This side of the figure is titled "Independent variable". The lowest of those four arrows is pointing at the box all the way on the right hand side. It has the caption "H1 direct effect". The box on the right is titled "Delegation performance" and is in the figure section with title "Dependent variable". The other three boxes are in the center with the title "Mediators". The first box is saying "Instance-Specific self-efficacy" and is connected to the other two boxes with arrows saying "H2 indirect effect". The second box is saying "Instance-Specific AI efficacy" and is connected to the other two boxes with arrows saying "H3 indirect effect". The second box is saying "Instance-Specific Difficulty Perception" and is connected to the other two boxes with arrows saying "H4 indirect effect".}
\end{figure}

We highlight our research model and the derived hypotheses in \Cref{research_model}. 
First, we answer the research question (\textbf{RQ \ref{rq1}}) on the impact of contextual information on human-AI team performance with the help of our research model.
We then test the hypotheses presented in this section to answer \textbf{RQ \ref{rq2}}.
In order to answer \textbf{RQ \ref{rq3}}, we perform subgroup analyses and explore how humans' perceived efficacy and humans' perceived difficulty correlate to the delegation behavior for varying types of contextual information.

\section{Methodology}
\label{methodology}

In this section, we outline the recruitment process of participants and the study design. We end this section by describing the development of the AI and the data we utilize for the study, along with the metrics we use to analyze the results.

\subsection{Data Selection}
\label{data_selection}

Similar to previous studies, we choose the task of estimating the incomes of US citizens \citep{bordt2022post, ezzeldin2023fairfed}. We use the dataset of \citet{ding2021retiring}. This dataset consists of ten features and one target variable and is based on the American Community Survey (ACS) Public Use Microdata Sample (PUMS). Overall, there are 1,599,229 datapoints in the dataset. As this work investigates the impact of additional data knowledge and AI knowledge on humans' delegation behavior, we choose this dataset since it represents a task domain that requires expertise for accurate predictions. In this context, access to additional information might lead to a performance increase. Thus, people who are unfamiliar with the task will have a lower performance in predicting income accurately compared to financial experts. At the same time, the dataset should also be not too difficult for humans to still enable the opportunity to bring in their unique knowledge in this human-AI collaboration \citep{fugener2021will}. 

To avoid cognitive overload through contextual information, we reduce the number of displayed features to four and validate it in our pre-test. Cognitive overload is a phenomenon investigated by previous research that leads to confusion and impairment in conducting a specific task \citep{paleja2021utility, dietzmann2022artificial}. The results of our pre-test show that participants, given four features, attained a performance of 57\%, which is similar to other studies that display all features to participants \cite{chen2023understanding}. Presenting these four features does not lead to cognitive overload.
To select the displayed features, we train a decision tree on a subset of the data and choose the four features with the largest feature importance.
The final four features used in the study are \textit{age}, \textit{education}, \textit{occupation}, and \textit{hours worked per week in the past 12 months}. We then randomly sample 60 instances of the holdout-set, of which participants are shown 12 random ones during the study.

\subsection{Study Design}
\label{study-design-section}

\begin{figure}[!b]
  \centering
  \includegraphics[width=\linewidth,clip]{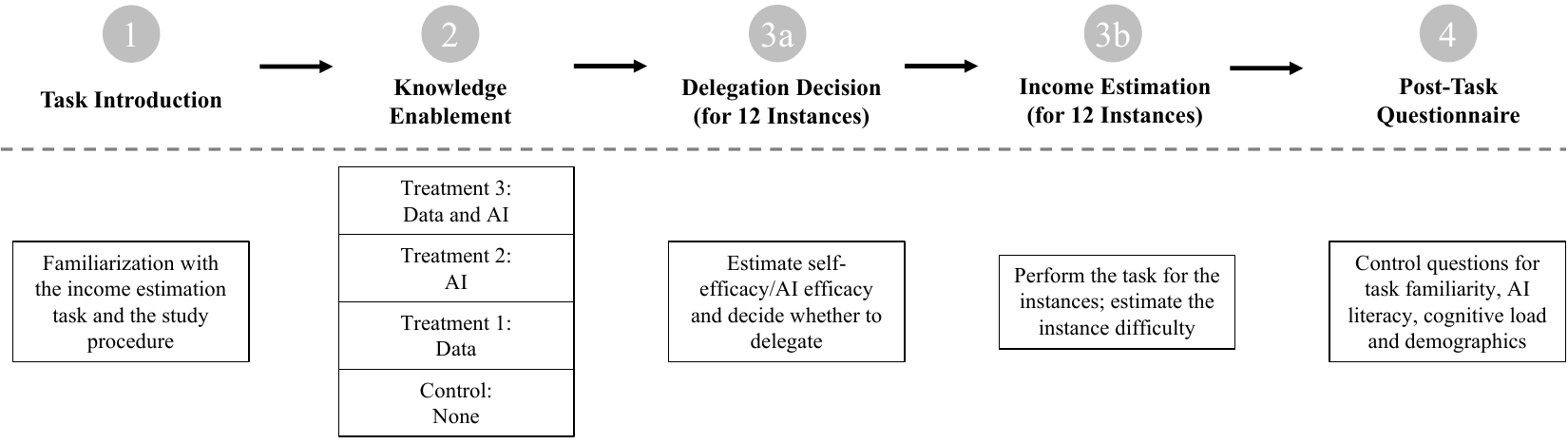}
  \caption{The design of the study is set up in four parts: first, participants are introduced to the task of the study, followed by part two in which they are randomly assigned to one of the four treatment groups. In part three, participants conduct the task of the study and have to fill out a questionnaire in part four.}
  \label{experimentdesign}
  \Description{The figure is showing the flow of the study. The words "1 Task Introduction", "2 Knowledge Enablement", "3a Delegation Decision", "3b Income Estimation" and "4 Post-Task Questionnaire" are connected with arrows from left to right. The following 5 descriptions belong to those captions: 1 - "Familiarization with the income estimation task and the study procedure", 2 - a table with Treatment 3: Data and AI, Treatment 2: AI, Treatment 1: Data and Control: None, 3a - Estimate self-efficacy/AI efficacy and decide whether to delegate, 3b - Perform the task for the instances, 4 - "Control questions for task familiarity, AI literacy, cognitive load and demographics".  
}
\end{figure}

We answer our research questions through a between-subjects experiment and investigate the effect of providing contextual information on human delegation behavior in an income prediction task. The IRB-approved study is set up in 4 phases. The study procedure follows the design outlined in \Cref{experimentdesign}, which we explain in further detail below.

Participants are familiarized with the task of estimating the income of US citizens and with the study procedure (part 1). In part 2, we assign participants randomly to one of four different groups: control group, in which participants are provided with no additional knowledge; data treatment group, in which participants are provided with additional knowledge on the dataset used in this study; AI treatment group in which participants are provided with additional knowledge on the AI and the data and AI treatment group in which participants are provided additional data and AI knowledge. The respective information provided to participants can be seen in \cref{sec: appendix} in \cref{fig:AI_knowledge} and \cref{fig:data_knowledge} and an example of the task interface in \cref{sec: appendix} in \cref{fig:Interface}.

In part 3, participants perform the task of estimating income. This part is divided into \textit{part 3a} and \textit{part 3b}. Overall, they are given twelve instances in a random order.
We do this to minimize the possibility of random effects caused by drawing instances that do not represent the overall dataset distribution.
For each instance, participants are asked in part 3a to rate their instance-specific self-efficacy and the instance-specific AI efficacy. To rate the instance-specific efficacies for humans themselves and the AI, participants have to indicate their valuation on a slider (ranging from ``0 \% (not at all)'' to ``100 \% (well suited)''). Then, they have to decide whether to delegate the instance to the AI or classify the instance themselves. For each instance, participants can access the information on the dataset and/or the AI in the respective treatments by clicking a button (see \cref{sec: appendix} \cref{fig:AI_knowledge} and \cref{fig:data_knowledge}). Inspired by previous research \citep{oppenheimer2009instructional, Lai2020}, we ensure that participants can first advance to the next instance after a minimum of ten seconds to make sure that they screen and consider all provided information. 

After considering their efficacies and deciding to delegate, participants have to conduct the classification task for all twelve instances. We design the study in this way to ensure no reflection mechanisms throughout task conduction for single instances that might impact the self-estimation on efficacy or the delegation behavior for following instances. Again, in part 3b, participants are presented with the samples of part 3a in random order. We ask them to classify the instance (``Do you think this person earns more than \$50,000 per year?'') and rate their perceived difficulty (``How difficult do you rate this instance?'') on a six-point Likert scale (ranging from ``very difficult'' to ``very easy'').

Finally, after evaluating the income for twelve instances, participants have to conduct a questionnaire (part 4) to assess their AI literacy, their familiarity with the task domain, their cognitive load, and their demographics (age, gender, employment, and education)

\subsection{Recruitment}
We recruit the participants (in total 240 -- 128 female, 110 male, and two not specified; average age: 39.4 years, SD: 12.76; median time: 20.42 min) through the platform Prolific.co. Previous research indicates that this platform is a reliable source of research data \citep{peer2017beyond, PALAN201822}. The sample includes participants from the United States. A screening mechanism is implemented on the Prolific platform. With the filter, we target individuals who are fluent in the English language. Our recruitment strategy is designed to involve participants without any further restrictions to gain a general understanding of our results and to be able to generalize our findings. Throughout the study, several attention checks are implemented to ensure only valid results as suggested by \citet{abbey2017attention}.

Participants who meet the stated criteria and complete the study's requirements receive a base payment of $2.25$\pounds. On average, participants take 19.8 minutes to complete the study. Additionally, participants are incentivized to conduct the task correctly by providing two bonuses: first, they receive five pennies per correct delegation decision, i.e., if they delegate and the AI is correct or they do not delegate and are correct themselves. Second, to ensure that participants maintain attention during their own classification even for instances they delegated, they will also receive an additional bonus of 5 pennies for each instance they classify correctly in part 3b of our study.

\subsection{Metrics}

Similar to previous work (e.g.,  \cite{hemmer2023human,fuegener2021exploring,taudien2022calibrating}), we assess participants' delegation behavior by investigating how often they delegate an instance to the AI. Moreover, we additionally measure participants' performance in classifying the instances themselves (human performance---measured on all instances independent of the delegated decision) and performing the task with the AI (collaboratively---human-AI team performance). The latter describes the scenario in which the performance is composed of correctly classified instances that the human decides to conduct themselves and correctly classified AI instances that the human decides to delegate to the AI. 

We are interested in how participants rate the efficacy on an instance level. To do so, we ask participants for each instance ``How well are you suited to solve this task?'' and ``How well is the AI suited to solve this task?''. Participants have to indicate their efficacy on a slider for which 0\% indicates ``not at all'' and 100\% indicates ``well suited''.

Next to the efficacy, we assess participants' perception of difficulty. \citet{steffel2018delegating} show in their study that participants delegate instances for which they perceive a high difficulty to intelligent agents. In order to explore the effect of the perception of instance difficulty \citep{spitzer2023perception}, participants rate their perceived instance difficulty after classifying the respective instance (on a six-point Likert scale ranging from ``very difficult'' to ``very easy''). Thus, we ask them: ``How difficult do you rate this instance?''. 

Finally, we establish several control variables in the study to investigate the underlying factors that influence the impact of contextual information on human-AI team performance. We control for participants' cognitive load as previous research suggests that the information in explanations displayed to humans can affect their decision-making behavior \cite{abdul2020cogam, hudon2021explainable, herm2023impact}. As we are presenting contextual information in the treatment groups, we measure participants' cognitive load on a five-point Likert scale by having them rate six validated items previous research has established \citep{klepsch2017development}. Furthermore, we assess participants' AI literacy by using the items of \citet{ehsan2021explainable} and participants' task familiarity by asking ``How familiar are you with the task domain of classifying peoples' income?'' on a six-point Likert scale (ranging from ``not familiar'' to ``very familiar'').
\section{Results}
\label{results_section}

In this work, we investigate the influence of contextual information on human-AI team performance when humans are able to delegate single instances to the AI. To do so, we conduct an empirical study with 240 participants. Through a between-subjects design, we analyze how different forms of contextual information affect the human-AI team performance and explore underlying human factors. In this section, we present the results of our study.

To address \textbf{RQ} \ref{rq1}, we perform thorough statistical tests in \Cref{sec: Impact of Performance} to compare the impact of different types of contextual information on human-AI team performance. In the following section, \Cref{sec: underlyfac}, we present the results of our mediation analysis, which follows the methodology of \citet{hayes2014statistical}. This analysis investigates the underlying human factors in the relationship between contextual information and human-AI team performance, based on the research model outlined in \cref{theoretical_section}. These findings help us answer \textbf{RQ} \ref{rq2}. Finally, we examine various human factors to determine their effects on delegation behavior using data from our study. We detail these results in \Cref{sec: delegbehav}, thus addressing \ref{rq3}.



\subsection{Impact of Contextual Information on Human-AI Team Performance}
\label{sec: Impact of Performance}

We now reveal the findings of the impact of contextual information on the human-AI team performance. As described earlier, we categorize the contextual information into two categories: data information (information about the feature's distribution concerning the target) and AI information (information about the AI's performance in regards to the features' distribution). 

\begin{figure}[!h]
    \centering
    \includegraphics[width=\linewidth]{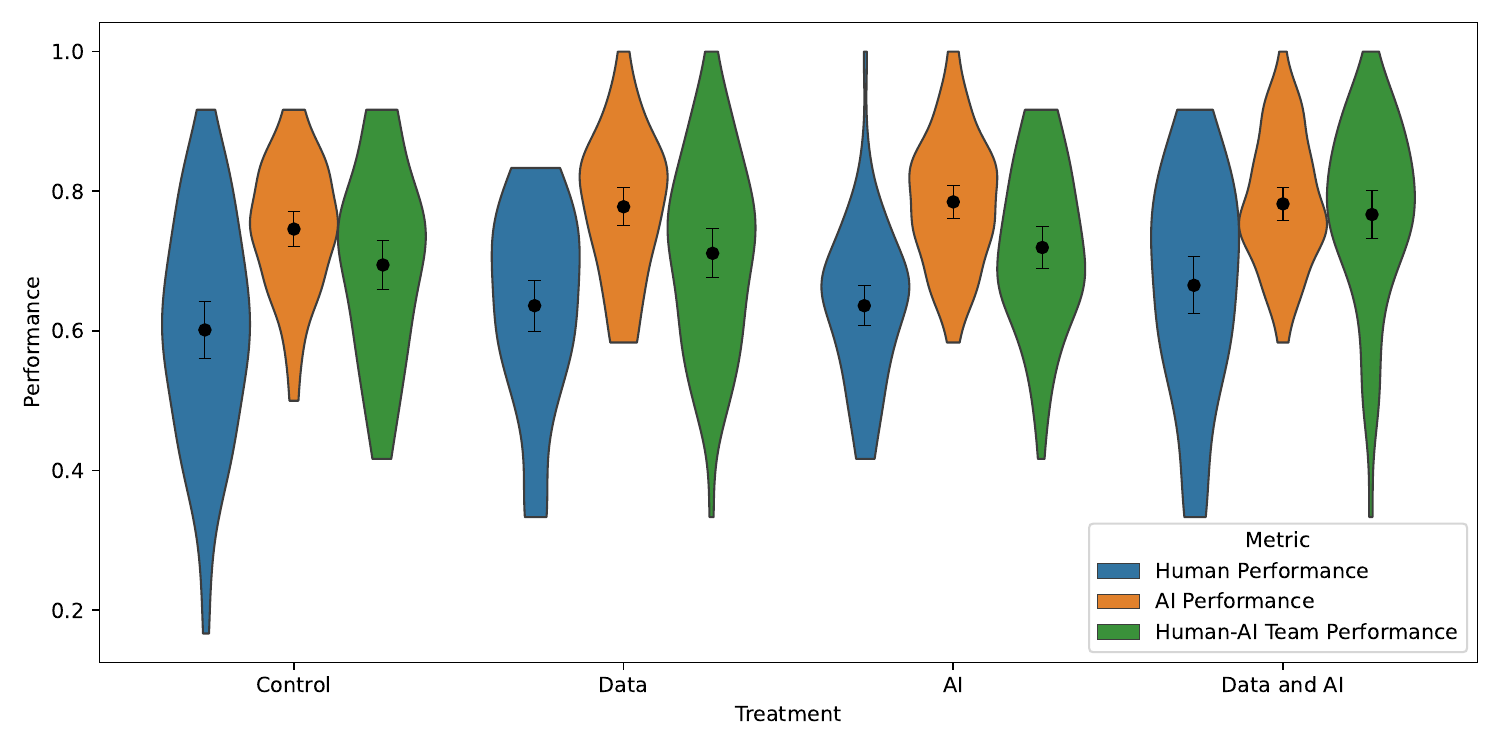}
    \caption{A comparison of the distribution of the mean performance per participant and 95\% confidence intervals across the four treatments: control treatment in which participants do not receive contextual information, data treatment in which participants receive contextual data information, AI treatment in which participants receive contextual AI information and the data and AI treatment in which participants receive contextual data and AI information.}
    \label{fig:performance comparison}
    \Description{A grouped violin chart is shown with three groups: "Mean Human Performance", "Mean AI Performance" and "Mean Human-AI Team Performance". The x-axis is titled "Treatment" with four labels: Control, Data, AI and Data and AI. The y-axis is titled "Performance" and is labeled from 0.2 to 1.0.}
\end{figure}

In our exploration of how contextual information influences the performance of human-AI teams, we calculate the performance on a per-participant basis across the twelve instances. \Cref{fig:performance comparison} shows an overview of the performances of human, AI, and human-AI team across the treatments.

To assess the effectiveness of combined human-AI efforts, we examine whether providing participants with additional contextual information can enhance the team performance. Specifically, we aim to examine if participants can delegate instances to the AI more effectively. First, we test our groups on equality of variances with Levene's test ($p = .7427$), indicating no statistical support for the presence of significantly different variances. Next, we use the Shapiro-Wilk test to check for normal distribution in our data. This reveals that three out of the four treatment groups do not follow a normal distribution. Given these findings, we proceed with the non-parametric Kruskal-Wallis test to investigate differences in the means of human-AI team performance. Given significant results ($H = 11.6348, p = 0.0087$), we perform a pairwise comparison to asses \textbf{RQ} \ref{rq1}. More specifically, we test whether the provision of contextual information increases the human-AI team performance compared to the control group. To do so, we use Mann-Whitney-U tests checking for increased performance and correct the p-values using the Holm-Bonferroni method to account for the multiple tests (see \Cref{tab: Tukey-Human-AI-Team}).

\begin{table}[h]
    \caption{Comparison of means for different treatment groups through Mann-Whitney-U test and subsequent correction of p-values for human-AI team performance.}
    \begin{threeparttable}
    \label{tab: Tukey-Human-AI-Team}
    \begin{tabular}{l l c c}
        \toprule
        Group 1 & Group 2 & Difference in means & Test statistic U\\
        \midrule
        Control & Data & .0167 & 1679\\
        Control & AI & .025 & 1633.5\\
        Control & Data and AI & .0722*** & 1215.5 \\
        \bottomrule
    \end{tabular}
    \begin{tablenotes}
        \item[1] Note: \textit{*} \textit{p < .1}; \textit{**} \textit{p < .05};  \textit{***} \textit{p < .01}\newline\newline
    \end{tablenotes}
    \end{threeparttable}
    \Description{A table is shown with three columns "Group1", "Group2" and "Difference in Mean". In the table notes the different p-value classifications are listed.}
\end{table}

The results indicate a difference in the performance of human-AI teams, with a significant mean difference ($p = .0087$). More specifically, the pairwise comparisons highlight that participants provided with both types of contextual information significantly outperform the control group ($p = .0026$). Conversely, teams receiving only one form of contextual information---either data treatment or AI treatment---do not show a significant improvement in performance ($p = .3738$) in both cases.
Given these outcomes, our research interest extends to understanding whether the provision of both types of contextual information can enhance human-AI team performance compared to participants with only one type of contextual information. To explore this, we replicate our approach using Mann-Whitney U tests and apply the Holm-Bonferroni method for p-value correction to maintain statistical rigor.
Here, we observe significant differences when both types of contextual information were provided as opposed to just one type. Specifically, participants with both data and AI contextual information improve their human-AI team performance compared to providing either contextual data ($U = 1342,p = .013$) or AI information ($U = 1336.5, p = .013$). By considering both types of contextual information, we answer \textbf{RQ} \ref{rq1}.
This highlights the synergetic potential of contextual information on data distribution and the capabilities of the AI. These findings emphasize the significance of providing thorough contextual information in optimizing human-AI collaboration. Our study shows that participants with comprehensive knowledge of data distribution and AI capabilities make delegation decisions more correctly, thereby enhancing the human-AI team performance. This indicates that there is a potentially additive effect when both types of contextual information are presented together, where capabilities of the AI and an understanding of data distribution appear to complement each other to optimize human-AI decision-making.

As the AI's performance exceeds the performance of the human participants in our study, a potential reason for this improved performance may lie in the number of delegated instances. Therefore, we next analyze the amount of instances delegated to the AI for the varying types of contextual information.

To analyze this, we conduct several statistical tests. First, we perform Levene's test to assess the homogeneity of variances ($p = .0154$), which indicates significantly different variances. Subsequently, we use the non-parametric Kruskal-Wallis test to compare medians across groups ($H = 10.9941, p = .012$), revealing significant differences in the number of delegated instances across treatments. Following this, we conduct a post-hoc Dunn test using the Holm-Bonferroni correction (see \Cref{tab: Dunn Delegation}), which shows that solely providing contextual data information results in significantly lower delegation compared to providing either contextual AI information alone ($p = .0113$) or providing both types of contextual information ($p = .0691$). An overview of the distribution of the number of delegated instances can be seen in \cref{sec: appendix} \cref{fig:delegation comparison}. Consequently, we observe that the increase in human-AI team performance is not solely due to a higher number of delegated instances to the AI, which has superior performance, but rather due to more effective delegation, i.e., delegating the instances for which the AI is correct.

\begin{table}[h]
    \caption{Comparison of means for different treatment groups through Dunn's test for the number of delegated instances.}
    \begin{threeparttable}

    \label{tab: Dunn Delegation}
    \begin{tabular}{l l c}
        \toprule
        Group1 & Group2 & Difference in means\\
        \midrule
        AI & Data and AI & -.2833 \\
        AI & Control & -.9667 \\
        AI & Data & -1.8167** \\
        Data and AI & Control & .6833 \\
        Data and AI & Data & -1.1533* \\
        Control & Data & -.85 \\
        \bottomrule
    \end{tabular}
        \begin{tablenotes}
        \item[1] Note: \textit{*} \textit{p < .1}; \textit{**} \textit{p < .05};  \textit{***} \textit{p < .01}\newline\newline
    \end{tablenotes}
        \end{threeparttable}
        \Description{A table is shown with three columns "Group1", "Group2" and "Difference in in mean of number of delegated instances". In the table notes the different p-value classifications are listed.}
\end{table}

To reveal whether access to the varying types of contextual information impacts participants' processing capacities, we examine the influence of contextual information on cognitive load.
Providing additional information may improve performance, but there is a risk of overwhelming individuals with details, which may negatively impact their performance.
However, our study finds that contextual information has a positive impact on human-AI team performances. To fully understand the impact of such information, we evaluate potential differences in cognitive load between treatments. Again, we first perform Levene's test to assess the equality of variances ($p = .2645$). The Shapiro-Wilk tests reveal that three out of four groups do not follow a normal distribution. Therefore, we perform the Kruskal-Wallis test ($H = 16.2446, p = .001$), which reveals significant differences in the means of cognitive load. Since we are interested in differences across all treatments, we use a post-hoc Dunn test and correct them using the Holm-Bonferroni method. As shown in \cref{tab: Dunn Cognitive Load}, significant differences were identified for both the data and AI treatment ($p = .0006$) and the AI treatment ($p = .0504$) compared to the control group. Further, we find that providing both types of contextual information also leads to increased cognitive load compared to the data treatment ($p = .0794$).
This suggests that contextual AI information is more difficult to process than contextual data information, thereby increasing cognitive load. However, even with this increased cognitive load, the integration of both types of contextual information significantly improves the performance of the human-AI team.

\begin{table}[h]
    \caption{Comparison of means for different treatment groups through Dunn's test for cognitive load.}
    \begin{threeparttable}

    \label{tab: Dunn Cognitive Load}
    \begin{tabular}{l l c}
        \toprule
        Group1 & Group2 & Difference in means\\
        \midrule
        AI & Data and AI & .0556 \\
        AI & Control & -.2361* \\
        AI & Data & -.125 \\
        Data and AI & Control & -.2917*** \\
        Data and AI & Data & -.1806* \\
        Control & Data & .1111 \\
        \bottomrule
    \end{tabular}
        \begin{tablenotes}
        \item[1] Note: \textit{*} \textit{p < .1}; \textit{**} \textit{p < .05};  \textit{***} \textit{p < .01}\newline\newline
    \end{tablenotes}
        \end{threeparttable}
        \Description{A table is shown with three columns "Group1", "Group2" and "Difference in in mean cognitive load". In the table notes the different p-value classifications are listed.}
\end{table}


\subsection{Underlying Factors in Human Delegation Performance}
\label{sec: underlyfac}

In the previous section, we show that the human-AI team performance differs for varying forms of contextual information. As we are interested in determining underlying human factors that influence this relationship, we conduct a thorough mediation analysis based on \Cref{theoretical_section} using data from all four treatments. This mediation analysis follows the Process Macro model of \citet{hayes2017introduction}. In this mediation analysis, we define contextual information as the independent variable and human-AI team performance as the dependent variable. We measure the effects on an instance basis, which is why human-AI team performance is modeled as binary variable\footnote{For this direct effects model, we report McFadden, CoxSnell, -2LL and ModelLL scores.}. As outlined in \Cref{theoretical_section}, we identify three possible underlying factors that might mediate the effect of contextual information on human-AI team performance: instance-specific self-efficacy, instance-specific AI efficacy, and instance-specific difficulty perception. We model these factors as mediator variables. As the independent variable contextual information is multi-categorical, we follow the protocol of \citet{hayes2014statistical} and code the four groups of contextual information through dummy variables (see \Cref{sec: appendix}, \Cref{coding_depv}). In this coding, we use the \textit{Control}-Group as the indicator group. This way, we can compare the effects of each treatment with the control group. Additionally, we model task familiarity, cognitive load, and AI literacy as control variables. We show the results of this mediation analysis in \Cref{mediationresults}.

\begin{table}[htbp!]
\caption{Mediation analysis of the influence of human factors on the relationship of contextual information on human-AI team performance.}
\begin{threeparttable}

\begin{tabular}{m{3.5cm} R{1.1cm} R{.5cm} R{.002cm} R{1.1cm} R{.5cm} R{.002cm} R{1.1cm} R{.5cm} R{.002cm} R{1.1cm} R{.5cm}} \hline
\multicolumn{1}{m{3.5cm}}{Dependant variable} & \multicolumn{2}{m{1.5cm}}{Team \newline performance} &&  \multicolumn{2}{c}{Self-efficacy} & & \multicolumn{2}{c}{AI efficacy} & & \multicolumn{2}{m{1.7cm}}{Difficulty \newline perception}   \\
\cmidrule{2-3} \cmidrule{5-6} \cmidrule{8-9} \cmidrule{11-12}
& coeff & se &  & coeff & se & & coeff & se &  & coeff & se \\
\hline \hline
Const   & 1.64***   & .36 & & .79*** & .02 & & .80***   & .02  && .43***  & .02     \\
Contextual information$:$ &&&&&&&&&&& \\
\textit{- None (baseline)}   &  & &  &  & &    &     \\
\textit{- Data}   & .05   & .12 & & .04*** & .01 & & .04***   & .01  && -.04***  & .01  \\
\textit{- AI}   & .07   & .12 & & -.01 & .01 & & .02**   & .01  & & -.02  & .01  \\
\textit{- Data and AI}   & .27**   & .13 & & .01 & .04 & & .07***   & .01  & & -.04***  & .01  \\
Self-efficacy   & -1.31***  & .24 &&   &  &&    &  &&   &   \\
AI efficacy   & 1.03***  & .27 &&   &  &&  & &&   &   \\
Difficulty perception   & -1.07***  & .19 &&   &  &&  & &&   &   \\
Task familiarity   & -.04  & .18 &&  .05*** & .02 && .01   & .01 && -.11***  & .02  \\
AI literacy   & -.13 & .20 &&  .03 & .02 &&  .09***   & .01  & & -.01  & .02   \\
Cognitive load   & -.08  & .37 &&  -.11*** & .03 && -.16***   & .03  && .10***  & .04   \\ \hline
McFadden / R   & \multicolumn{2}{c}{.02} & & \multicolumn{2}{c}{.15} & & \multicolumn{2}{c}{.21}  & & \multicolumn{2}{c}{.14}  \\
CoxSnell / $R^{2}$   & \multicolumn{2}{c}{.02} &&  \multicolumn{2}{c}{.02} & & \multicolumn{2}{c}{.05}  & & \multicolumn{2}{c} {.02}     \\
-2LL /MSE   & \multicolumn{2}{c}{333.51} &&  \multicolumn{2}{c}{.04} & & \multicolumn{2}{c}{.03}  & & \multicolumn{2}{c}{.06}   \\ 
ModelLL / F   & \multicolumn{2}{c}{68.91***} &&  \multicolumn{2}{c}{11.51***} & & \multicolumn{2}{c}{23.64***}  & & \multicolumn{2}{c}{9.86***}   \\\hline
\end{tabular}
    \begin{tablenotes}
        \item[1] Note: \textit{*} \textit{p < .1}; \textit{**} \textit{p < .05};  \textit{***} \textit{p < .01}\newline\newline
    \end{tablenotes}
    \end{threeparttable}

\begin{tabular}{m{2cm} m{.6cm} m{.6cm} m{.6cm} m{.6cm} m{.002cm} m{.6cm} m{.6cm} m{.6cm} m{.6cm} m{.002cm} m{.6cm} m{.6cm} m{.6cm} m{.6cm}} \hline
& \multicolumn{14}{c}{Relative indirect effects}\\
\cmidrule{2-15}
& \multicolumn{4}{c}{Self-efficacy} &&  \multicolumn{4}{c}{AI efficacy} & & \multicolumn{4}{c}{Difficulty Perception}  \\
\cmidrule{2-5} \cmidrule{7-10} \cmidrule{12-15}
& Effect & Boot SE & Boot LLCI & Boot ULCI & & Effect & Boot SE & Boot LLCI & Boot ULCI & & Effect & Boot SE & Boot LLCI & Boot ULCI \\
\hline \hline
Contextual information$:$ &&&&&&&&&&& \\
\multicolumn{2}{m{2.95cm}}{\textit{- None (baseline)}}   &  & &  &  & &    &     \\
\textit{- Data}   & .06   & .02 & -.09  & -.02 && .04   & .01 & .02  & .07 && .04   & .02 & .02  & .08 \\
\textit{- AI}   & .02   & .16 & -.01  & .05 && .02   & .01 & .00  & .04 && .02   & .01 & -.01  & .05 \\
\textit{- Data \& AI}   & .01   & .02 & -.03  & .04 && .06   & .02 & .03  & .11 && .04   & .02 & .02  & .08  \\ \hline
\end{tabular}

\label{mediationresults}
\Description{Two tables are shown. In the higher table, there are 9 columns: dependent variable, Delegation Performance - coeff, Delegation Performance - se, Self-efficacy - coeff, Self-efficacy - se, AI efficacy - coeff, AI efficacy - se, Difficulty Perception - coeff, Difficulty Perception - se. In the table notes the different p-value classifications are listed. In the lower table there are 13 columns. The firs colum represents the variables of the regression analysis, followed by the four metrics "Effect", "Boot SE", "Boot LLCI" and "Boot ULCI" for the three mediators "Self-efficacy", "AI efficacy" and "Difficulty Perception".}
\end{table}

The data shows that the effect of providing contextual information on the human-AI team performance is mediated by participants' self-efficacy, AI efficacy, and difficulty perception of an instance. To analyze each mediation, we draw from \citet{zhao2010reconsidering} to classify each effect and outline the mediations' implications.

We start this analysis by focusing on the effects of \textbf{contextual data information} compared to the baseline for which no contextual information is provided. \Cref{mediationresults} indicates that the indirect effect of contextual information on the human-AI team performance is significant for self-efficacy, AI efficacy, and difficulty perception with indirect effects of $.06$, $.04$, and $.04$, respectively. As the direct effect of contextual information on human-AI team performance is not significant, we classify these effects as an indirect mediation \citep{zhao2010reconsidering}.
This suggests that providing participants with contextual data information not only increases their confidence in their ability to classify the instances themselves (indicated by an increase in self-efficacy coefficient, $coeff_\text{self-efficacy}=.04$) but also enhances their confidence in the AI's capability to accurately classify the instances (as shown by an increase in AI efficacy coefficient, $coeff_\text{AI efficacy}=.04$).
Moreover, having access to this kind of contextual information decreases the difficulty perception of single instances ($coeff_\text{difficulty perception}=-.04$). Based on the contextual data information, participants achieve a higher human-AI team performance. Thus, our findings do not support hypothesis \ref{hyp1} but support hypotheses \ref{hyp2}, \ref{hyp3}, and \ref{hyp4} for this form of contextual information.

Next, we analyze the effects of \textbf{contextual information based on AI}. The data reveals a significant indirect effect for AI efficacy as a mediator with a size of $.02$. The direct effect of contextual information on the human-AI team performance is not significant. Thus, we categorize this effect as indirect mediation. Participants increase their confidence in judging the AI as being suitable to conduct the instance with access to this contextual data ($coeff_\text{AI efficacy}=.02$). Through this knowledge, participants increase their human-AI team performance. Thus, they can judge when to delegate an instance more accurately. Hence, our results do not support hypotheses \ref{hyp1}, \ref{hyp2}, and \ref{hyp4} but support hypothesis \ref{hyp3} for this form of contextual information.

Lastly, we explore the data on \textbf{contextual information based on both} forms of additional context. Overall, we see a significant direct effect for AI efficacy and difficulty perception. With the direct effect of contextual knowledge on participants' human-AI team performance being significant, we classify the effect of AI efficacy as complementary mediation (positive direct effect of AI efficacy on human-AI team performance) and the effect of difficulty perception as competitive mediation (negative direct effect of difficulty perception on human-AI team performance) \citep{zhao2010reconsidering}. Having access to both sources of contextual knowledge increases participants' confidence in judging the AI as suitable for classifying the instance and, at the same time, decreases the difficulty perception for single instances ($coeff_\text{AI efficacy}=.07, coeff_\text{difficulty perception}=-.04$).
Overall, participants are more likely to delegate the right instances and are more likely to classify instances themselves correctly. Thus, the findings do not support hypothesis \ref{hyp2} but support hypotheses \ref{hyp1}, \ref{hyp3} and \ref{hyp4}.

\subsection{Impact of Perceived Efficacy and Difficulty on Delegation Behaviour}
\label{sec: delegbehav}
Understanding the effects of instance-specific efficacy and perceived task difficulty is essential to assessing delegation behavior. Delegation decisions are not always based solely on objective measures, as subjective perceptions and self-assessments can significantly affect decision-making \cite{pinski2023ai}. To derive insights into the impact of human factors on human-AI delegation scenarios for varying types of contextual information, we specifically analyze how these factors impact human delegation behavior.

Therefore, we examine the influence of human factors on the delegation behavior---specifically, whether an instance was delegated or not---for varying types of contextual information. To do this, we analyze each delegation decision and the associated human factors individually, allowing us to understand the impact of individual decisions. Specifically, we seek to understand how subtle variations in human factors might influence a participant's decision to delegate. This granularity provides a richer perspective on the interplay between human perceptions and the specifics of the instance at hand. To achieve this, we run a logit regression with contextual information, perceived difficulty, self-efficacy, and AI efficacy as independent variables. The dependent variable is the delegation behavior (see \cref{sec: appendix} \cref{tab:logit-results}).

The regression model is significant ($p = .000$). At a more granular level, each human factor is also significantly associated with the delegation decision. The positive coefficient for difficulty (1.195, $p = .000$) implies that as perceived difficulty increases, so does the likelihood of delegating the instance to the AI. This suggests that participants are more willing to delegate instances to the AI when they find them challenging.
Similarly, AI efficacy has a positive effect with a coefficient of 7.864 ($p = .000$). This means that when participants perceive the AI as more effective, they are more likely to delegate instances to it. This underscores the importance of perceived AI capabilities in influencing humans to delegate instances. In contrast, self-efficacy has a negative effect with a coefficient of -9.7956 ($p = .000$), meaning that when humans judge themselves as suitable for an instance, they are less likely to delegate.
Among the treatments, contextual AI information is significant ($p = .026$) with a coefficient of .306, suggesting that when participants are informed about the capabilities of an AI, they are more likely to delegate. 
In contrast, contextual data information, while weakly significant ($p = .055$), has a negative coefficient of $-.261$. This suggests that when participants are given this particular type of contextual information, they are less likely to delegate instances to the AI. Consistent with previous observations, this shows that having insight into the data distribution makes participants feel more equipped to handle instances themselves. Interestingly, the data and AI treatment has no significant effect ($p = .994$).

\section{Discussion}
\label{discussion_section}

Next, we discuss our results from \cref{results_section} to answer our research questions. Further, we discuss the broader implications of our findings on the design of human-AI collaboration scenarios. 

\textbf{A holistic approach to information provision is imperative for enhanced performance}.
Our study demonstrates that the combination of contextual information on data distribution and AI performance significantly improves human-AI team performance (see \Cref{sec: Impact of Performance}). While providing both types of information simultaneously enhances human-AI team performance, offering only one type does not yield in significant improvements. 
In another study \cite{pinski2023ai}, the authors show that contextual information on the AI, in the form of AI literacy, improves human-AI delegation. Similarly, prior work finds that providing contextual information about the AI (e.g., the AI's certainty score for a prediction \cite{taudien2022calibrating}) can improve performance and users' trust in the AI. We extend this knowledge and show that the provision of both types leads to the best results in terms of human-AI team performance.
Participants demonstrate greater proficiency in delegating tasks to the AI when they have insightful information about both the underlying data distribution and the AI's predictive performance. Consequently, the performance of the human-AI team is synergistically enhanced.
However, it remains unclear why providing only a single type of contextual information fails to improve human-AI team performance. A potential reason is that successful human-AI collaboration requires understanding the complementary capabilities of both the human and the AI \cite{hemmer2024complementarityhumanaicollaborationconcept}. Providing a single type of contextual information supports the assessment of either the human’s (via data distribution) or the AI’s (via AI performance) capabilities. Thus, offering both types of information enables the human delegator to comprehensively assess the capabilities of both parties, leading to more effective task delegation and increased team performance.
Thus, our findings suggest that improving human-AI team performance requires a comprehensive approach to information provision. This involves understanding the data distribution and gaining insights into the AI’s capabilities to assess both the human delegator and the AI in correctly solving a task. This comprehensive approach ultimately facilitates effective task delegation to the more suitable team member.

In line with previous research in HCI, which demonstrates that additional information, such as behavior descriptions of an AI, can enhance human-AI team performance \citep{cabrera2023improving}, we stress the importance of a holistic approach to contextual information that includes both the AI and the underlying data. This finding extends previous work, which provides only a single type of contextual information on the AI, resulting in improved human-AI team performance. For example, \citep{pinski2023ai} show that contextual information on the AI, in the form of AI literacy, improves human-AI delegation. Similarly, \cite{taudien2022calibrating} demonstrate that providing contextual information about the AI, such as the AI’s certainty score for a prediction, can improve performance and users’ trust in the AI.

\textbf{Providing contextual AI information increases cognitive load.}
Prior work shows that contextual AI information leads humans to align their delegation behavior with their AI efficacy and self-efficacy but that this type of information does not positively impact the delegation performance \cite{pinski2023ai}.
Interestingly, while our findings show that participants face challenges in capturing and processing information about the AI's predictive performance, expressed through an increased cognitive load (see \Cref{sec: Impact of Performance}), this does not correspond to a lower performance. As users process rich information about the AI's predictive performance, they risk overlooking or misinterpreting critical details of the instances themselves. This could lead to impaired delegation decisions, especially in scenarios where a thorough understanding of the AI's strengths and weaknesses is critical. The more cognitive resources devoted to understanding the AI, the fewer resources are available for decision-making tasks. However, in our study, participants maintained or even improved their performance despite these potential drawbacks.

Zooming in, we find that providing participants with contextual AI information increases cognitive load, whereas providing them with contextual data information does not. This suggests that contextual AI information is more complex for participants to process compared to contextual data information. One potential explanation for this discrepancy is that contextual data information involves a straightforward thought process---it only requires understanding the impact of a specific attribute on the target variable. In contrast, contextual AI information indicates the likelihood of an AI being correct without providing an explanation of the underlying reasoning. This lack of explanation may prompt participants to engage with and examine the AI’s assessments in depth to ascertain whether the AI is correctly evaluating the attributes.
Nevertheless, despite this potential engagement with the underlying relationships between specific attributes and the target variable, we do not observe an improvement in performance when only contextual AI information is provided. Improved performance is only observed when both types of contextual information are combined.
Nevertheless, these findings underscore the potential of well-designed contextual information and highlight the need for an intuitive representation of the AI's predictive abilities, ensuring users are informed without feeling overwhelmed.
\citet{hudon2021explainable} make similar findings that additional information about the AI's prediction increases the cognitive load. We extend this observation in the context of delegation tasks and, additionally, find that contextual information on the underlying data seems to trigger less cognitive load, which shows that humans have more difficulty in processing information about the AI than the data.

\textbf{Self-efficacy, AI efficacy, and difficulty perception mediate the effects of contextual information on human-AI team performance.}
Through our mediation analysis, we reveal underlying human factors that influence human-AI team performance. Similar to previous studies  \citep{hemmer2023human, pinski2023ai, lubars2019ask}, we show that humans' self-efficacy, AI efficacy, and difficulty perception at the instance-level mediate the effects of contextual information on delegation behavior (see \Cref{sec: underlyfac}). \citet{hemmer2023human} provide evidence that higher self-efficacy leads to better performance in human-AI collaboration. Our data shows similar results for a human-AI delegation scenario.

Looking more closely at our results for different types of contextual information and comparing the effects across treatments, we see different levels of impact on human factors. Specifically, we find that displaying contextual data information significantly increases instance-specific self-efficacy and AI efficacy while significantly decreasing perceived difficulty. In contrast, for contextual AI information, only the increase in perceived AI efficacy is significant. 
A potential reason for this disparity is that showing the distribution of data relative to the target variable helps participants understand the task better and identify characteristics relevant to whether a person is earning more than \$50,000. This understanding may enhance their confidence in their ability to solve the task successfully, which is reflected in their increased self-efficacy. Conversely, showing only the AI's accuracy concerning specific attributes and the target variable helps participants assess whether the AI's predictions are correct but not which attributes lead to higher income. Consequently, their self-efficacy is not affected by this assessment, although they attribute greater reliability to the AI in successfully predicting whether a person is earning more than \$50,000.
When comparing the treatments with both types of contextual information to those with only a single type, we observe that the combined effect of both types of contextual information on AI efficacy is amplified, with a coefficient of .07 compared to .04 for data information alone and .02 for AI information alone. This suggests a reinforcing effect between both types of contextual information. Contextual data information may help participants identify attributes associated with higher income, and this intuition may then be validated by contextual AI information. For example, if participants recognize that individuals working more than 60 hours per week often earn more than \$50,000, and the AI accurately assesses this attribute, participants may perceive the AI as more reliable in solving specific instances.

\textbf{Humans' delegation behavior is influenced by their contextual knowledge.} Recent research clusters different types of human delegation behavior and finds that not all humans make use of the option to delegate tasks to an AI \cite{fuegener2021exploring}. We advance this view and analyze the delegation behavior for different types of contextual information. Beyond the contextual information on data distribution, our research highlights the essential importance of gaining insights into the AI's performance (see \Cref{sec: Impact of Performance}). While understanding data distribution provides contextual knowledge, gaining insights into the AI's capabilities is important as well. By knowing the AI's performance, users can calibrate their decisions about when to delegate tasks and when to take over \cite{taudien2022effect}, leading to improved human-AI team performance. Our findings suggest that the real potential of contextual information is realized when these two elements---data distribution and insights into AI performance---are combined.

\subsection{Implications of Contextual Information on CSCW Settings}
\label{sec: Implications}

Next, we discuss the implications of our findings on the design of CSCW systems for human-AI collaboration, more specifically, human delegation.

\textbf{Contextual information can serve as a means to compensate for missing expertise.} Our study highlights the important role of contextual information about data distribution in the collaboration of humans and AI. For example, it could serve as a valuable substitute for individuals without extensive domain knowledge, offering insights and patterns that would typically require years of experience. Conversely, having contextual data information is an additional layer of information for those who already have domain expertise that potentially complements and enhances their existing understanding.
However, the effectiveness of this strategy depends on balancing the presentation of this information. When addressing non-experts, it is imperative to ensure that data distribution information is accessible and not overwhelming \cite{abdul2020cogam} to avoid cognitive overload. For domain experts, the challenge is to integrate this information in a way that augments rather than conflicts with their inherent domain understanding. Ultimately, the goal is to facilitate a harmonious synergy between human intuition---whether it stems from domain expertise or insights from data distribution---and AI analysis. This approach offers a promising path for more effective and informed decision-making in human-AI collaborations. It emphasizes both the depth of expert knowledge and the breadth of data insights, making it a valuable asset in any collaborative work setting.

\textbf{Designing effective CSCW systems for human delegation to AI necessitates careful balancing of different forms of contextual information.}
Our findings indicate that the type of contextual information provided to participants influences their delegation behavior. Specifically, providing only contextual data information reduces the tendency to delegate tasks, leading participants to solve tasks themselves more often than when provided with contextual AI information or both types of contextual information. In contrast, providing solely contextual AI information increases the tendency to delegate but does not enhance human-AI team performance.
These insights have practical implications for designing human-AI collaboration systems in workplace settings as they offer valuable guidance on manipulating delegation dynamics through strategic selection of contextual information. For instance, if the goal is to encourage employees to solve tasks independently, supplying only contextual data information may be effective. However, this approach relies on the human delegator's ability to solve tasks accurately and to identify the instances for which the AI is correct. 
One practical application of these findings could be during the onboarding of new employees. Initially, contextual data information can help supplement the lack of expertise, aiding employees in developing domain knowledge. As their expertise grows, introducing contextual AI information can encourage more delegation to the AI, thereby enhancing the overall efficiency of human-AI collaboration.

\subsection{Limitations and Future Work}
\label{sec: Limitations}
While our study provides valuable insights into human delegation to AI systems through the provision of specific information, there are several limitations that need to be acknowledged.

One limitation of the study arises from task specificity. The findings, which were derived from an income prediction task utilizing the US Census dataset, may not generalize to other tasks with varying difficulties and domains. Consequently, broadening the scope to include a broader range of tasks in subsequent research would refine our understanding of the impact of contextual information on delegation decisions in different applications.

In our study, we use bar graphs to present contextual information to participants. While this approach effectively communicates basic data trends and AI performance metrics, it may have limitations in communicating more complex aspects of AI behavior and data characteristics. Future research should consider the use of more nuanced visualization techniques, which could potentially increase the depth of understanding in human-AI collaboration. These advanced methods could offer a more comprehensive view of AI decision-making processes, aiding users in making more informed delegation choices. Investigating the effectiveness of such visualization enhancements in conveying complex AI-related data remains a valuable avenue for future research, for example, complementing the delegation task with interactive data exploration methods.

Our study does not reach the benchmark of complementary team performance~\cite{hemmer2021human,bansal2021does,green2019principles}, where the collective capabilities of human-AI collaborations exceed the performance of each entity operating independently. An interesting direction for future research involves integrating explainable AI mechanisms into the delegation decision-making process. By offering explanations, users might be assisted in determining the AI's ability to accurately classify an instance \cite{taudien2022calibrating} and thereby refine their choices. Examining the intersection of AI-generated explanations and the provision of contextual information presents a promising direction.

Another limitation arises from the nature of the data utilized in our study. Our method for supplying contextual information is designed for structured data, which inherently displays observable patterns and quantifiable attributes. The applicability of this approach may differ significantly when used on unstructured data such as images or text. Thus, future research should investigate effective methodologies for conveying contextual information for unstructured data sets. Expanding our paradigm to include these diverse data types will allow us to determine the robustness and adaptability of our findings and potentially identify innovative strategies for enhancing human-AI collaboration across broader data landscapes.

\section{Conclusion}
\label{conclusion_section}

Providing access to the right form of contextual information is crucial for computer-supported and collaborative work. In this study, we set out a research design to explore how different forms of contextual information affect human-AI collaborations. In doing so, we take a human-centered perspective and investigate the underlying human factors that influence human decision-making in collaborations where humans can delegate individual instances to an AI. To do so, we establish a theoretical understanding and develop a research model with which we empirically validate our hypotheses. We show that the provision of contextual information increases the human-AI team performance. This effect is highest when humans get access to both information on the AI performance and on the data distribution. We also reveal that humans' self-efficacy, AI efficacy, and difficulty perception for single instances impact their delegation behavior. Overall, we contribute to CSCW and HCI with a thorough understanding of how to improve human-AI collaboration systems when humans can delegate instances to an AI and provide insights into underlying mechanisms that influence such scenarios. 

In summary, our study enhances our understanding of human-AI decision-making by emphasizing the importance of contextual information in human-AI interaction. Our results offer instructions to practitioners on how to customize collaborative systems to meet individual needs and direct researchers toward promising avenues of future research on effective human-AI decision-making. As the field continues to evolve, the findings and approaches presented in this study provide a solid foundation for further exploration and innovation. Extensive and rigorous research is needed to fully understand and exploit the role of contextual information in human delegation. We invite researchers to join this debate and hope to inspire scientists to participate actively in this endeavor.

\bibliographystyle{ACM-Reference-Format}
\bibliography{references}

\appendix
\newpage
\section{Appendix}
\label{sec: appendix}

\begin{figure}[!h]
    \centering
    \includegraphics[width=\linewidth]{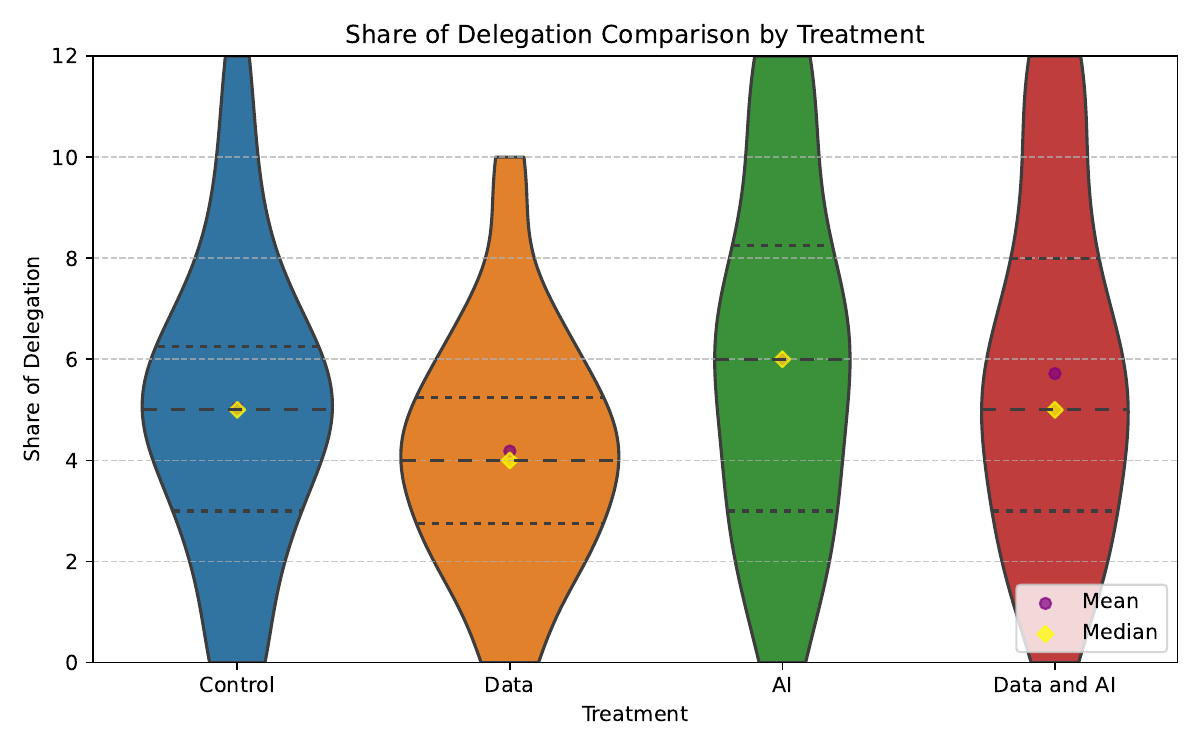}
    \caption{A comparison of the distribution of the mean of the number of delegated instances per participant and 95\% confidence intervals across the four treatments.}
    \label{fig:delegation comparison}
    \Description{A grouped violin chart is shown that displays the mean number of delegated instances per treatment. The x-axis is titled "Treatment" with four labels: Control, Data, AI and Data and AI. The y-axis is titled "Share of Delegation" and is labeled from 0 to 12.}
\end{figure}

\begin{table}[htbp!]

\caption{Coding of the dependent variable \textit{contextual information} in the mediation analysis}

\begin{tabular}{P{4cm} P{1cm} P{1cm} P{1cm}}

\hline

Contextual information &  X1 & X2 & X3 \\ \hline \hline

\textit{None} & 0 & 0 & 0 \\  \hline
\textit{Data} & 1 & 0 & 0 \\  \hline
\textit{AI} & 0 & 1 & 0 \\  \hline
\textit{Data \& AI} & 0 & 0 & 1 \\  \hline
\\\\ \end{tabular}

\label{coding_depv}
\Description{A table with four columns is shown. The first column says "Contextual Information", the second column "X1". the third column "X2", the fourth column "X3". In the first column, the rows are stating "None", "Data", "AI" and "Data and AI".}
\end{table}

\begin{table}[h]
\centering
\caption{Logit Regression Results}
\label{tab:logit-results}
\begin{tabular}{@{}lllllcc@{}}
\toprule
\multicolumn{7}{c}{\textbf{Model Summary}} \\
\midrule
\textbf{No. Observations} & 2880 && \textbf{Df Model} & 6 && \\
\textbf{Df Residuals} & 2873 && \textbf{Pseudo R-squ.} & 0.3129 && \\
\textbf{Log-Likelihood} & -1355.3 && \textbf{LL-Null} & -1972.7 && \\
\textbf{LLR p-value} & 1.478e-263 && & & \\
\midrule
\textbf{Variable} & \textbf{Coef.} & \textbf{Std. Err.} & \textbf{z} & \textbf{P$>$|z|} & \textbf{[0.025} & \textbf{0.975]} \\ 
\midrule
Difficulty & 1.1950 & 0.211 & 5.670 & 0.000 & 0.782 & 1.608 \\
Self efficacy & -9.7956 & 0.426 & -23.000 & 0.000 & -10.630 & -8.961 \\
AI efficacy & 7.8643 & 0.478 & 16.464 & 0.000 & 6.928 & 8.801 \\
AI information & 0.3058 & 0.138 & 2.223 & 0.026 & 0.036 & 0.576 \\
Data information & -0.2610 & 0.136 & -1.920 & 0.055 & -0.527 & 0.005 \\
Data and AI information & -0.0011 & 0.135 & -0.008 & 0.994 & -0.267 & 0.264 \\
Intercept & 0.4343 & 0.322 & 1.347 & 0.178 & -0.198 & 1.066 \\
\bottomrule
\end{tabular}
\Description{A table that describes the details of the performed regression analysis on the delegation behaviour. In the top part, a summary of the model is provided and in the bottom a more detailed version of the coefficients and power are reported.}
\end{table}

\begin{figure}[htbp!]
    \centering
    \includegraphics[width=0.75\linewidth]{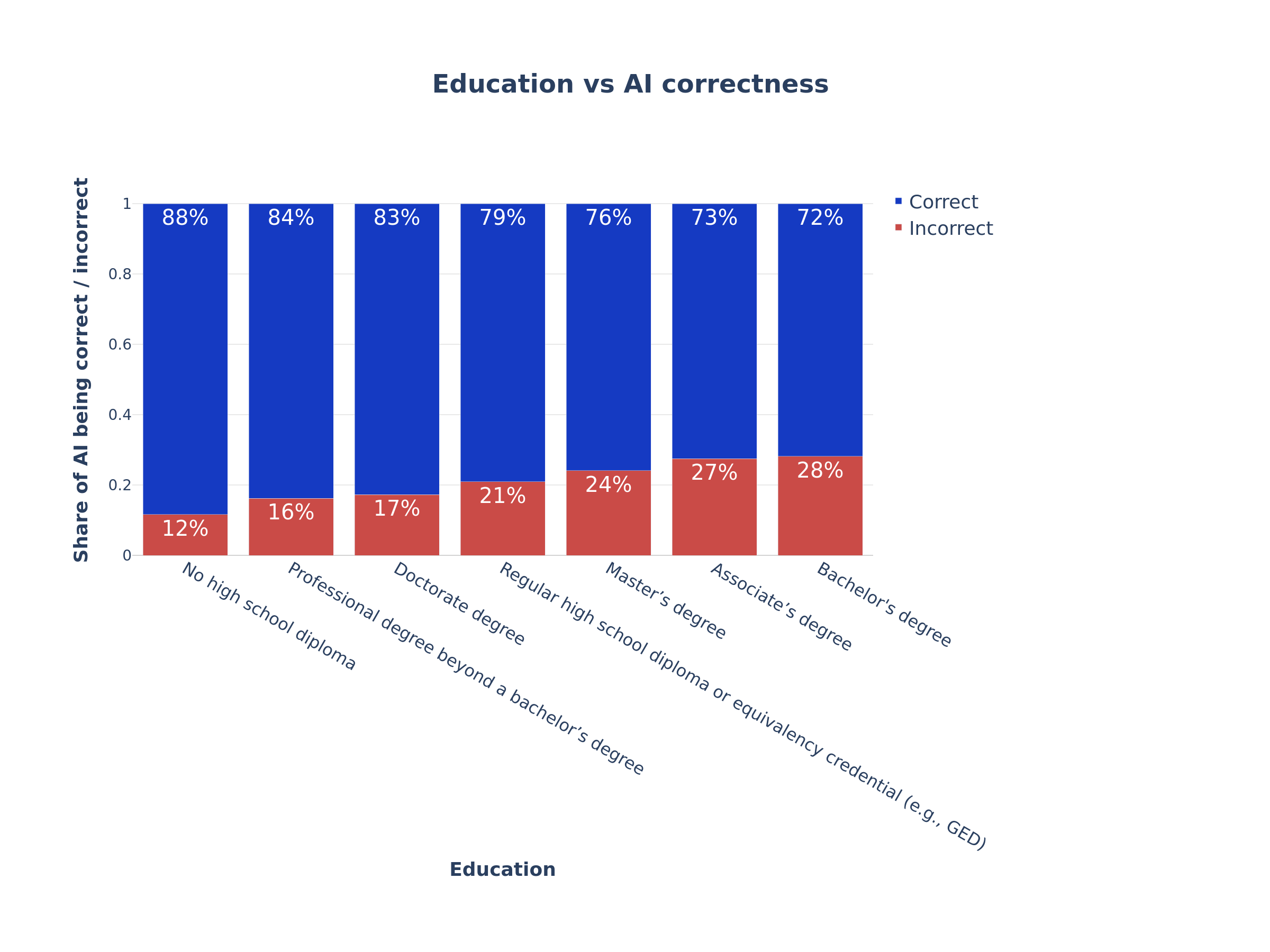}
    \caption{Provided contextual AI information for the feature education}
    \label{fig:AI_knowledge}
    \Description{A grouped bar chart is shown with two groups: "Correct" and "Incorrect". The x-axis is titled "Education" with labels No high school diploma, Professional degree beyond a bachelor's degree, Doctorate degree, Regular high school diploma or equivalency credential (e.g., GED), Master's degree, Associate's degree and Bachelor's degree. The y-axis is titled "Share of AI being correct / incorrect" and is labeled from 0 to 1.0.}
\end{figure}

\begin{figure}[htbp!]
    \centering
    \includegraphics[width=0.75\linewidth]{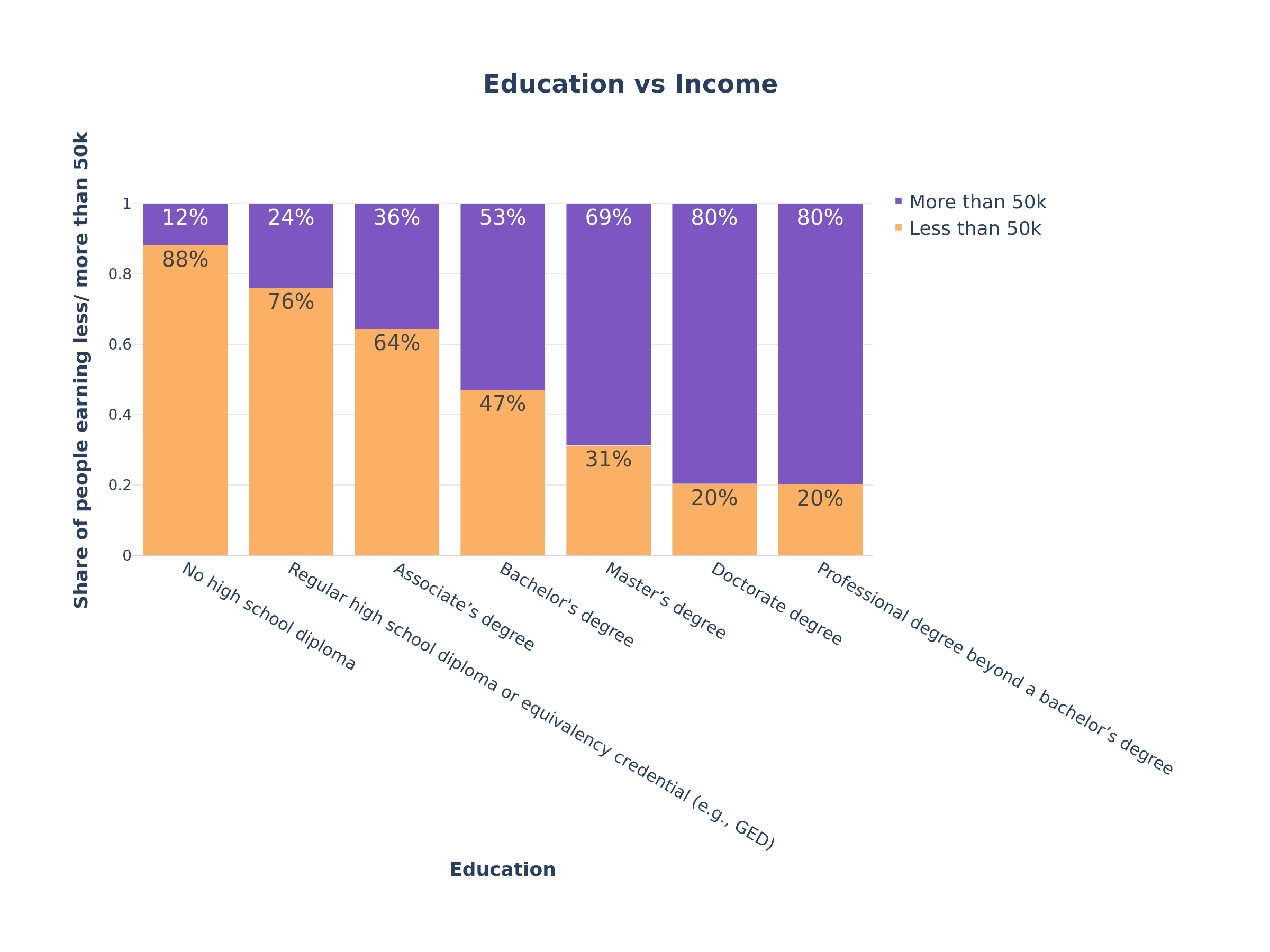}
    \caption{Provided contextual data information for the feature education}
    \label{fig:data_knowledge}
        \Description{A grouped bar chart is shown with two groups: "More than 50k" and "Less than 50k". The x-axis is titled "Education" with labels No high school diploma, Professional degree beyond a bachelor's degree, Doctorate degree, Regular high school diploma or equivalency credential (e.g., GED), Master's degree, Associate's degree and Bachelor's degree. The y-axis is titled "Share of people earning less / more than 50k" and is labeled from 0 to 1.0.}
\end{figure}

\begin{figure}[htbp!]
    \centering
    \includegraphics[width=0.75\linewidth]{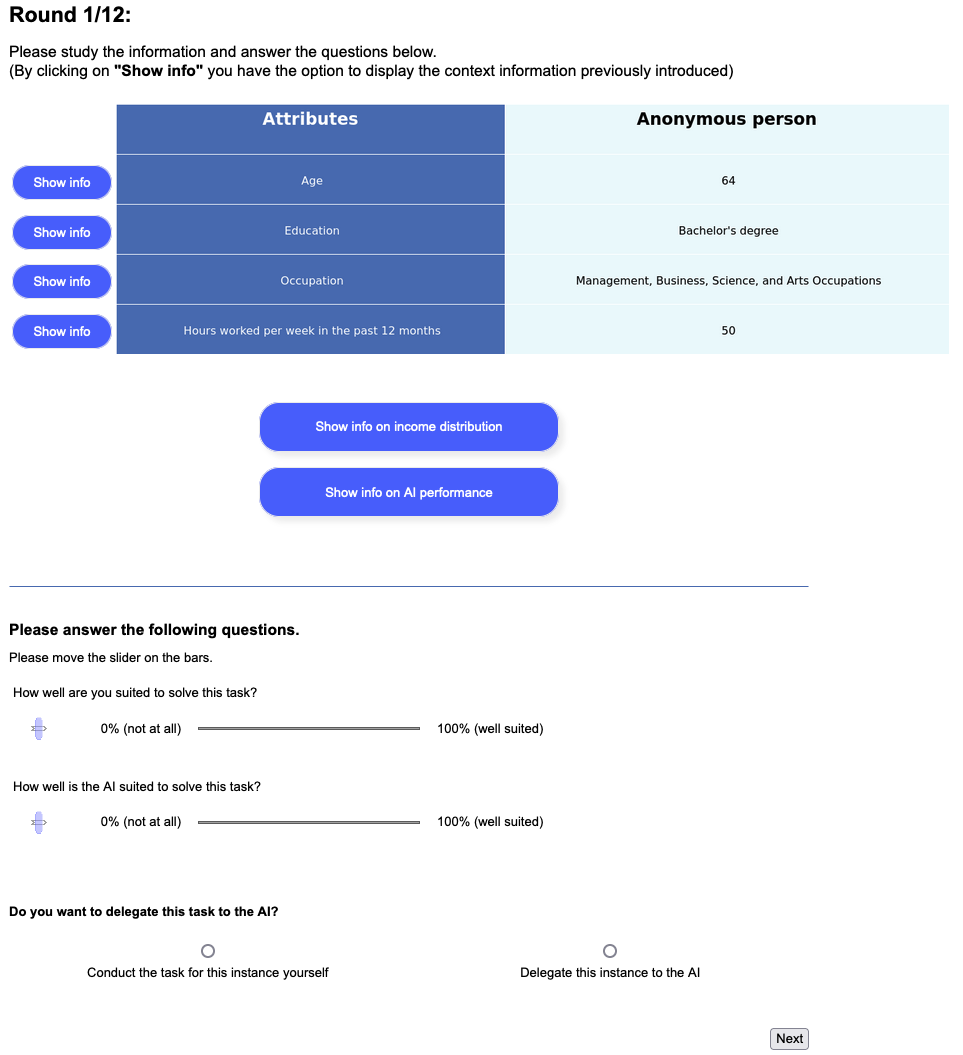}
    \caption{Provided interface for participants in the \textit{Data and AI} treatment}
    \label{fig:Interface}
        \Description{A table with the feature information of the specific instances. On the left are buttons to access the data information and below the table are information on the overall distribution of the income and AI performance. Below these buttons are the task questions. First, two sliders ranging from 0 (not at all) to 100 (well suited) to asses self-efficacy and AI-efficacy.  Finally, there is the question whether the participant want to delegate this task to the AI or not and a button to go the next instance.}
\end{figure}

\end{document}